# Unidirectional Luminescence from Quantum Well Metasurfaces

Prasad P. Iyer*[1], Ryan A. DeCrescent*[2], Nikita A. Butakov[1], Abdullah Alhassan[3,4], Guillaume Lheureux[3,4], Claude Weisbuch[3,4], Shuji Nakamura[1,3,4], Steven P. DenBaars[1,3,4] and Jon. A. Schuller[1].

1. Department of Electrical and Computer Engineering, University of California Santa Barbara
2. Department of Physics, University of California Santa Barbara
3. Department of Material Science and Engineering, University of California Santa Barbara
4. Solid State Lighting and Energy Electronics Center, University of California Santa Barbara

**III-Nitride light emitting diodes (LEDs) are the backbone of ubiquitous lighting and display applications. Imparting directional emission is an essential requirement for many LED implementations. Although optical packaging[1], nano-patterning[2,3] and surface roughening[4] techniques can enhance LED extraction, directing the emitted light requires bulky optical components. Optical metasurfaces provide precise control over transmitted and reflected waveforms, suggesting a new route for directing light emission. However, it is difficult to adapt metasurface concepts for incoherent light emission, due to the lack of a phase-locking incident wave. In this Letter, we demonstrate metasurface-based design of InGaN/GaN quantum-well structures that generate narrow, unidirectional transmission and emission lobes at arbitrary engineered angles. We show that the directions and polarization of emission differ significantly from transmission, in agreement with an analytical Local Density of Optical States (LDOS) model. The results presented in this Letter open a new paradigm for exploiting metasurface functionality in light emitting devices.**

Light emitting diodes (LED) are rapidly enabling solid-state solutions to commercial lighting applications. Imparting unidirectionality to LEDs is a challenging, problem with a host of applications[5–7] awaiting a scalable solution. Directional emission is naturally observed in lasing systems,[13,14] where all of the resonators are emitting coherently. By modifying the local density of optical states (LDOS) through coupling to e.g optical nanoantennas[8–11], researchers can impart directivity on ensembles of incoherently emitting local point sources. However, requirements for precise positioning and challenges of achieving high emitter densities preclude scaling to large areas.[12] At infrared frequencies, directional thermal emission has similarly been achieved through coupling to nanoantennas[15] and gratings[16,17]. Moving beyond these initial demonstrations, greater control over emission directionality may be possible using phased array metasurface concepts[18–20]. For instance, phased-array lensed thermal emission has recently been theoretically demonstrated[17,21]. Such directed thermal emitters typically exploit a peaked LDOS and losses associated with surface waves. Absorption losses are necessary for thermal emitters, but constrain the applicability to bulk visible frequency LEDs. Spatially uniform metasurfaces have been used to enhance and redirect PL, However, as yet there have been no demonstrations of directed PL enabled through coupling to metasurface-defined spatially varying phase profiles. Here, we experimentally demonstrate asymmetric and unidirectional visible frequency luminescence from light-emitting phased array metasurfaces.

Traditional metasurfaces transform a coherent wave by imprinting sub-wavelength spatial phase profile across the surface through resonant interaction with plasmonic or dielectric resonators[20,22–25]. Consider a transmitting metasurface (at a wavelength, λ = 540nm), designed to redirect an incident plane wave (at $\theta_i$) to an output plane wave (at $\theta_t$) according to the grating equation, $\boldsymbol{k}_0 \sin \theta_t = \boldsymbol{k}_0 \sin \theta_i + \boldsymbol{k}_M$. Here, $\boldsymbol{k}_M = \nabla \phi$ is the desired in-plane "metasurface momentum" to be imparted to a transmitting wave and $k_0=2\pi/\lambda$ is the free-space momentum of light with wavelength λ. Our passive transmitting metasurface (Fig. 1A) is designed to be compatible with light-emitting InGaN/GaN multiple QW structure grown by metal-oxide chemical vapor deposition (MOCVD) techniques upon a double-side polished sapphire substrate (Methods). InGaN QWs are grown at a height, $h_{QW}$=1.35 µm above the substrate where the GaN defect density is significantly reduced, followed by a 100 nm GaN capping layer, yielding a total film thickness of h=1.45 µm. We directly structure this film into arrays (periodicity, p=250nm) of high-aspect ratio (h/w=5.5-8.5) nanopillars (Fig. 1B; Methods). The metasurface phase, ϕ(x), is controlled by varying the nanopillar widths. However, rather than simulate the transmission phase of an incident beam, we simulate the accumulated phase of light emitted into the substrate from in-plane oriented point dipoles located 100 nm from the top of a GaN nanopillars. The FDTD simulations assume that the point dipole sources are radiating in-phase with respect to each other, such that they only emit along the surface normal, in analogy with typical transmission/reflection simulations that only consider normal incidence illumination. The numerically calculated emission phase and normalized amplitude are shown in Fig 1C; 0 to 2π phase control is achieved by varying nanopillar width between 150-210 nm. With this emission phase, we map the required spatial phase profile for beam deflection (based on the grating equation) to the expected nanopillar width at each point on the metasurface. We construct and demonstrate metasurfaces to deflect normal incidence ($\theta_i$=0) light into desired transmission lobes within the sapphire substrate ($\theta_t$ between 10°-80° or equivalently $\mathbf{k}_t/k_0$ between 0.17 to 0.98) and subsequently measure the PL from InGaN QWs embedded within the metasurface nanopillars.

High-efficiency, polarization-independent unidirectional metasurface transmission lobes are measured via energy-momentum spectroscopy[26–30] (Methods) and shown in Figures 2A and 2B. The momentum-resolved transmission of normal incidence 540nm light is plotted as a function of the normalized in-plane electromagnetic wave momentum, $k_{||}/k_0$. Accordingly, the circle $|k_{||}/k_0|=1$ forms the light-cone of free-space emission modes, outside of which modes are trapped within the substrate ($n_{sub}=1.77$; $1<\frac{|k|}{k_0}<n_{sub}$) or the GaN thin film ($n_{GaN}\approx2.2$; $n_{sub}<\frac{|k|}{k_0}<n_{GaN}$). For both s- (Fig 2A) and p- (Fig 2B) polarized light, the normal-incidence beam is deflected into a very narrow unidirectional transmitted lobe at the design momentum $k_M$ ($\theta_t$).

It is difficult to predict how the metasurface patterning will affect overall luminescence because the emitting QWs naturally emit over a broad angular range (Supplementary Fig. 3). Energy-momentum spectra of photoluminescence from the optically pumped (Methods) metasurfaces are shown in Figures 2C and 2D. The s-polarized PL (Fig. 2C) shows no clear signatures of directional emission, and very little variation between different metasurface designs. On the other hand, the p-polarized PL is predominantly emitted into a single unidirectional lobe defined by the in-plane momenta $|k_{||}|= -k_0+k_M$ ($\theta_e=-90°+\theta_t$) (Fig. D). Consequently, as we increase $|k_M|$ to steer the transmitted beam to larger angles, we see that a strong and unidirectional *p*-polarized PL lobe is translated toward **k**=0 (Fig 2D). Unlike uniform nanopillar arrays[31], we break the inherent symmetry of the emission pattern by introducing a linear phase gradient. The emission is highly directed with a full width at half max (FWHM) less than ±5° in the substrate. These results demonstrate the promise for redirecting PL into desired metasurface channels, but the observed strong polarization dependence highlights the complexity of this phenomenon and the need to better understand its origins.

To better understand the observed emission properties, we develop a simple model starting from analytical calculations of the momentum-dependent emission (the LDOS) from an unstructured film. P-polarized light emission from the unstructured film (Supplementary information S3) is "naturally" concentrated in emission lobes just beyond the critical angle ($|\mathbf{k}|=\pm k_0$) that ordinarily are trapped within the substrate. These relatively sharp p-polarized emission maxima are a consequence of the approximate $\lambda/2n_{GaN}$ (100 nm) distance of the emitter from the GaN-air interface (Supp. Section S1). These intrinsically narrow, but symmetric, emission lobes in turn produce even narrower and asymmetric emission lobes after interacting with the metasurface structure. To understand this phenomena, we consider a simple model that accounts for both the metasurface *micro*periodicity $G=2\pi/p$ (p=250nm is the inter-pillar spacing) and *macro*periodicity P, which depends on the $2\pi$ phase wrapping distance and varies between each metasurface. This microperiodicity along the $\hat{x}$ and $\hat{y}$ directions defines a momentum, $G_0=2\pi/p$, that couples wave vectors $\mathbf{k}_{||}$ to the harmonics $\{\mathbf{k}_{||}\pm G_0\hat{x}\}$ and $\{\mathbf{k}_{||}\pm G_0\hat{y}\}$[3]; a new LDOS thus accounts for this microperiodicity by summing the original LDOS over a grid of points in *k*-space separated by $\pm G_0$ along both $\hat{x}$ and $\hat{y}$ axes (Supplementary Fig. 1B). The ratio $G_0/k_0\approx 2.15$ makes the bright thin-film emission lobes between neighboring zones strongly overlap along the $k_x$-axis. We then shift the resultant summed LDOS by an asymmetric component, $\mathbf{k}_M=+2\pi/P\,\hat{x}$ along the direction of the phase gradient (Supp. Sect. S3).

A comparison of experimental far-field PL images (left panels) with our analytical LDOS model (right panels) is made in Figure 3. The PL is collected with a polarizer oriented such that line cuts along $k_x=0$ ($k_y=0$) are purely p-(s-) polarized. Our simple model very well captures the location, shape, and width of the observed directional emission lobes across the five displayed metasurfaces. This excellent agreement between calculations and experiments validates the basic premises of our analytical model, providing insight into how to design future metasurface-modified light emitters. In particular, these results highlight the importance of starting with a highly directional LDOS prior to patterning, and considering the effects of both micro- and macro-periodicities.

We quantify the "beam efficiencies" for GaN metasurfaces operating in both PL (Fig 4A) and transmission (Fig. 4B) modes. Beam efficiency is defined as the fraction of light intensity in the main lobe to the total light collected in the far-field. In transmission, the metasurfaces demonstrate greater than 80% efficiency for small deflection angles with the efficiency dropping close to 50% for near-grazing transmission. In photoluminescence, the metasurfaces similarly show higher efficiency for near-normal emission while the beam efficiency reduces for near-grazing emission angles. This efficiency fall-off has been previously studied in other transmitting gratings and metasurfaces[32,33] suggesting that more complex unit-cell designs may be leveraged for higher efficiencies at steeper angles.

Encouragingly, compared to the thin film, our emitting metasurfaces show an increase in total PL intensity while also directing the emission in a dramatically narrower (1/18) lobe into the sapphire substrate. Structuring the emitting layer into uniform nanopillar arrays yields a 30-fold increase in the collected PL, concomitant with a substantial blue-shift (~160 meV) of the peak emission to 540 nm (Fig. 4B; solid colored lines). A strain induced piezo-electric field in the InGaN QWs relaxes during the formation (dry etching) of the nanopillars, leading to a larger recombination energy and increased electron-hole (e-h) wave function overlap in the QWs (Supp. Sect. S5). The increased PL intensity from the nanopillar array is attributed to increased absorption at the pump (405nm) wavelength (Supp. Sect. S1) and improved e-h wave function overlap due to a reduced quantum-confined stark effect. For example, in the extreme case with a nominal nanopillar diameter of 265nm and pitch 250nm, the system can be described as an array of holes in a GaN thin film; we observe only a modest blue shift in the PL wavelength (10nm) and a slight increase (10%) in the PL intensity (Fig. 4B). In contrast, for smaller widths, free-standing nanopillars are completely strain relaxed, blue shifting the emission wavelength to 540nm. These results demonstrate that dry etched InGaN/GaN nanopillar arrays increase the luminescent output at green wavelengths. Metasurface patterning, then, provides two major benefits for future LED technologies: 1) increasing PL quantum yield through strain relaxation and 2) redirecting the emitted light into preferred directions, focused spots, or other engineered waveforms.

The results presented in the Letter demonstrate that metasurface design methodologies may be used to enhance and engineer the luminescence from thin films. We demonstrate that phased array metasurfaces with strategically embedded QW emitting layers can be designed to emit p-polarized light in a unidirectional manner, breaking the inherent symmetry conditions of traditional PCs. Fundamentally, we have demonstrated that the same design principles used to direct phase-coherent incident sources can used to direct incoherent light emission through LDOS engineering. The results presented in this letter point the way to new classes of metasurface-based light emitters where enhanced and directional emission is attained without external packaging components.

**Methods:**

**Growth and Fabrication:** The sample was grown hetero-epitaxially on (0001) c-plane double side polished sapphire substrate by atmospheric pressure metal-organic chemical vaper deposition (MOCVD). The sample structure consisted of an AlGaN nucleation layer grown at

570°C. A 1 µm unintentionally doped (UID) GaN template layer grown at 1200°C. The active region consisted of three periods multiple quantum wells (MQWs) with a 3 nm InGaN QW, a 2 nm $Al_{0.3}Ga_{0.7}N$ cap and a 10 nm GaN barrier. Finally, a 100 nm UID GaN protective buffer is grown on top which not only protects the emitting QW from fabrication damage but also provides the necessary boundary conditions for the emitted light to be channeled through the etched nano-pillar into the substrate. Electron beam lithography is done to generate the required metasurface patterns using Hydrogen silsesquioxane (60nm thick) as the resist on a double hard-mask layers of Ruthenium (30nm) and $SiO_2$ (400nm) on the GaN thin film. The hard masks where dry etched using an inductively coupled plasma (ICP) etch using $Cl_2/O_2$ (at 100W power) -gas mixture for Ru and $CHF_3/CF_4$ gas mixture (at 50W power) for $SiO_2$. Following the oxide etch, the Ru metal was dry etched away before GaN etch in a reactive ion chamber. An initial $BCl_3$ etch at 100W for 2 minutes was performed on the sample to remove the native oxide on the GaN (to prevent micro masking) followed by a pure $Cl_2$ etch at 100W power. This etch was developed to achieve extremely high-aspect ratio GaN nanopillars (>40:1) with a vertical side-wall (Fig 2A) to etch through the 1.25µm layer and stop on the sapphire substrate The residual oxide hard mask was removed using Buffered HF dip for 60s.

**FDTD Simulations:** Numerical calculations were performed in Lumerical FDTD. For absorption enhancement calculations, a uniform pillar array with a 250 nm periodicity was simulated. Pillars (refractive index of 2.23, total height 1450 nm) contained three absorbing layers (3 nm thick) representing the quantum wells at heights of 1300nm, 1315nm, and 1330. Mesh sizes were kept at 4nm×4nm×6nm in the bulk of the pillar and reduced to 4nm×4nm×0.5nm in the absorbing layers. To estimate Purcell enhancements, we simulated large regions of phased metasurface gratings with in-plane oriented electric dipole sources at each unique dipole location within macro-periodic super cell of the GaN metasurface.

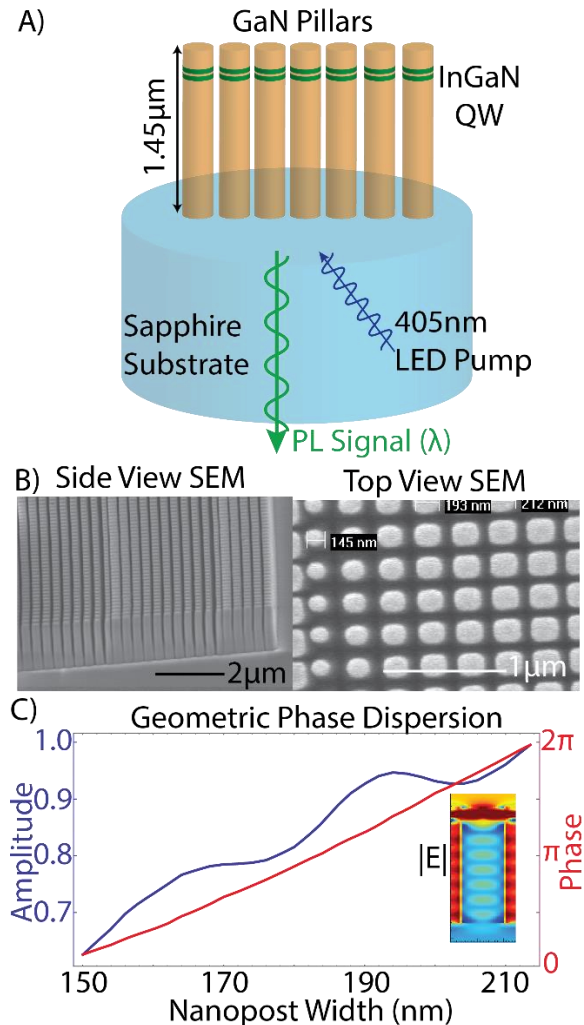

**Figure 1: A)** Sketch of the GaN nanopillars with InGaN MQWs grown on a double side polished sapphire substrate which is pumped from below using a 405nm LED and emitted PL is collected below the substrate. **B)** SEM images of the GaN metasurface formed showing the high aspect ratio resonators formed with a well-defined size gradient only along the x-direction (p-pol). **C)** The phase (red, right axis) and normalized amplitude (blue, left axis) of the light emitted from the point in-plane (x,y) electric dipole embedded in the GaN nanopillar into the substrate as a function of the square nanopillar width. The inset shows the electric field profile of the excited mode by the point dipole with the red (blue) representing maximum (minimum) in field intensity. **D)** The field plots shows the resonator coupling in the s-(top panel) and p- (bottom panel) polarized emission from a 3 resonator super cell. The plots shows that the only p-pol emission is sensitive to the phase gradient of the metasurface.

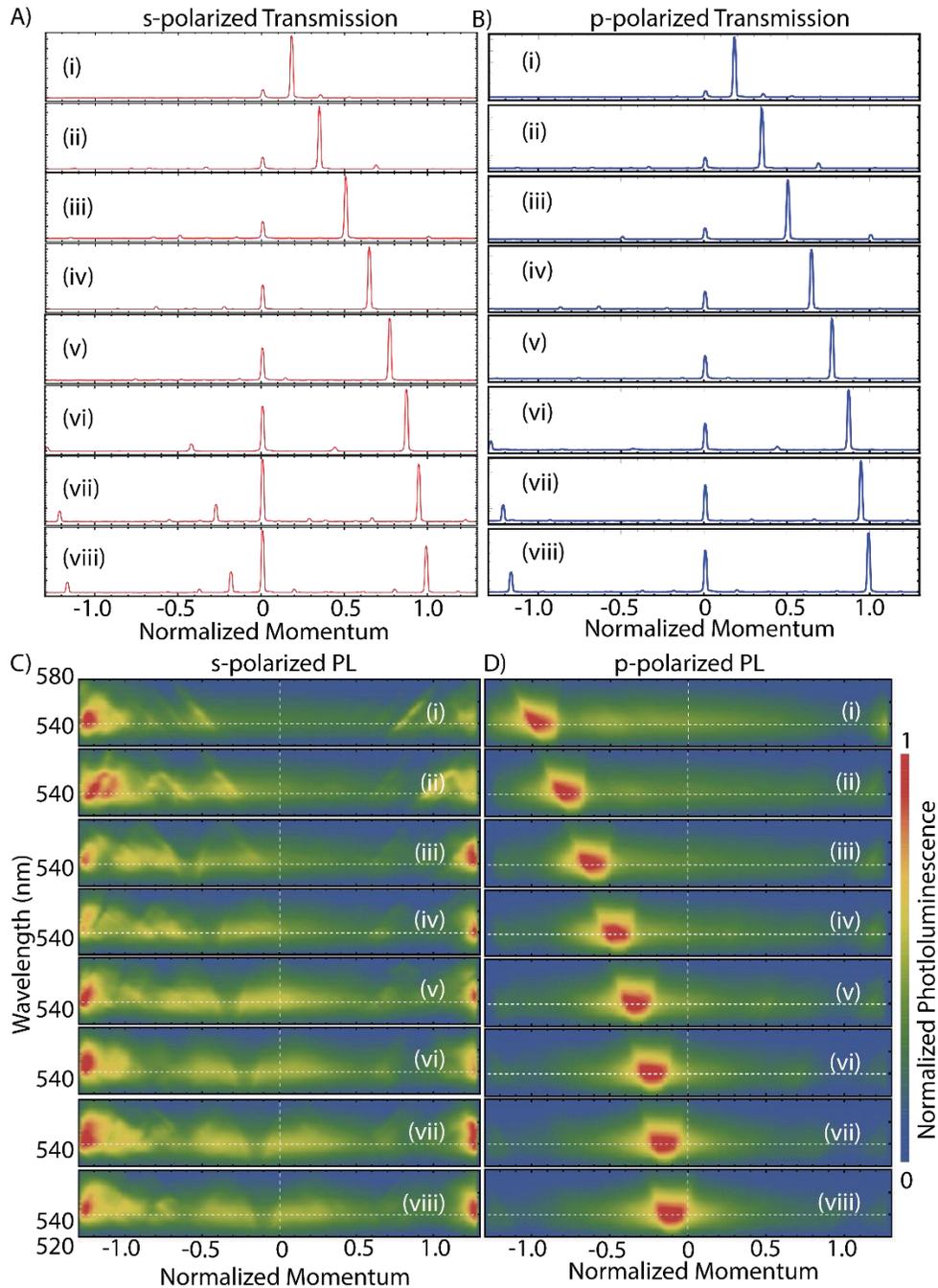

**Figure 2: A and B)** S-polarized and P-polarized transmission through the metasurface is plotted as a function of the normalized in-plane momentum for the same 8 (a-h) metasurface from panel A. These metasurfaces were designed to deflect the normally incident beam (0°) in transmission to 10°(a)-80°(h) in steps of 10° using a linear-phase profile. The specular transmission at 0° shows the relative in-efficiency of the metasurface. **C and D)** Normalized s-polarized and p-polarized PL intensity is plotted as a function of wavelength (y axis, 520-580nm) and normalized in-plane momentum ($k_{||}$) for 8 different metasurfaces in each panel (a-h). The p-pol PL peak is shifted in momentum space from -1 to 0 based on the metasurface design.

There is a dip in the s-pol PL intensity for s-pol light corresponding to the peak in the p-pol PL intensity plots.

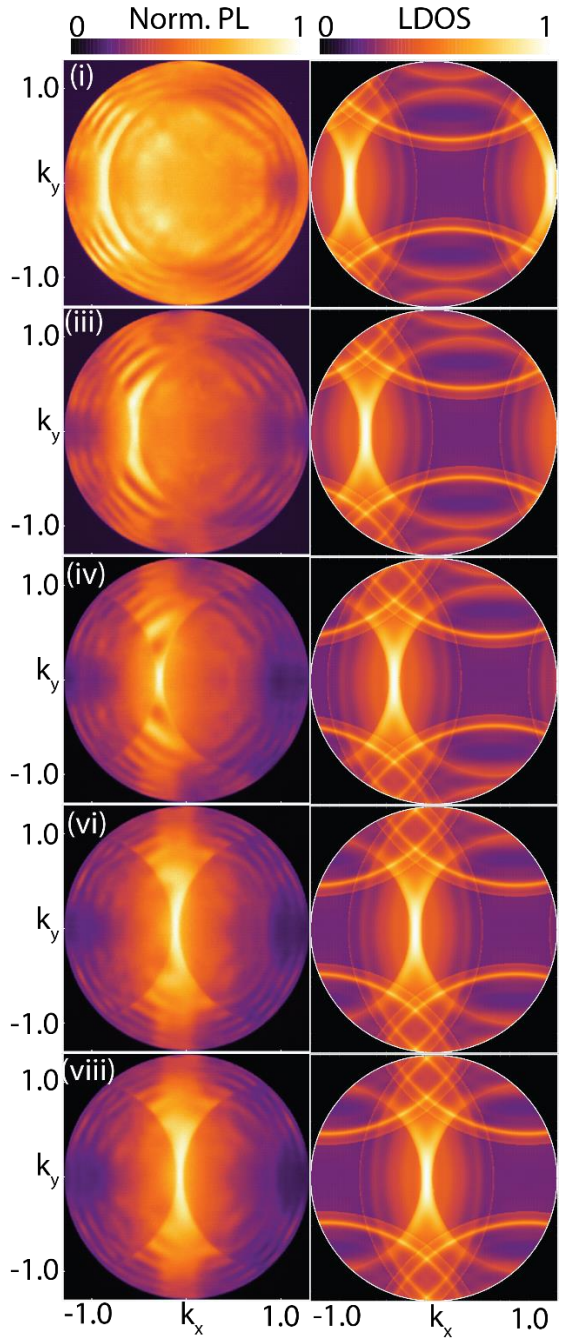

**Figure 3. A)** Intensity-normalized experimental far-field p-polarized PL as function of normalized momenta $k_x/k_0$ and $k_y/k_0$. A vertical polarizer has been applied such that light along $k_x=0$ ($k_y=0$) is p-(s-) polarized. Each panel label (i-viii) corresponds to a metasurface design in Fig 2. The strong PL peak at $k_x=1$ (a) shifts toward $k_x=0$ (viii) while no shift is observed in $k_y$. **B)** The modified LDOS based on the photonic crystal effects and the metasurface-induced $k_x$

translation. The circular features and the intense asymmetric emission maximum at $k_x=0$ match very well with the experimental data panels in Fig 4A.

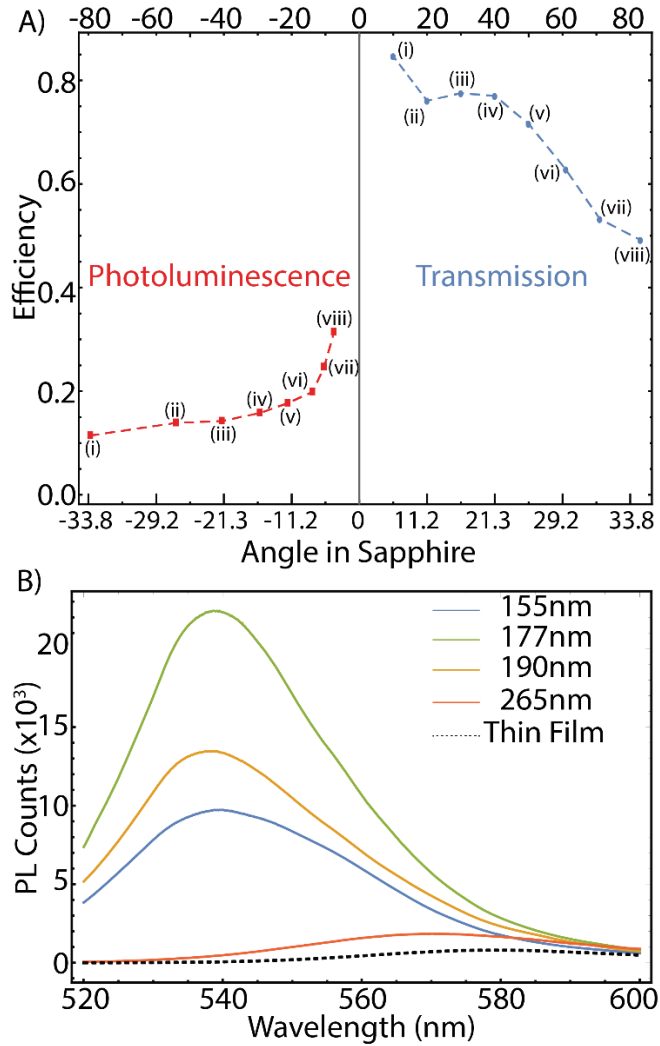

**Figure 4. A)** The efficiency (power in the main lobe/ total power) of transmission (right, blue) and PL (left, red) are plotted as function of angle in the sapphire substrate (bottom axis) and angle in air (top axis). As the metasurface deflects or emits away from the normal direction, the efficiency decreases for both transmission and PL. Every point on the curve corresponds to a designed metasurface plotted in Figure 2. **B)** The measurement setup used to collect momentum (angle) resolved PL at the back focal plane (BFP) using a high NA (=1.3) oil immersion lens. The blue color path indicates the 405nm pump LED while the green color path shows the PL collection scheme on the spectrometer.

# Supplementary Information

## Unidirectional Luminescence from Quantum Well Metasurfaces

Prasad P. Iyer*[1], Ryan A. DeCrescent*[2], Nikita A. Butakov[1], Abdullah Alhassan[3,4], Guillaume Lheureux[3,4], Claude Weisbuch[3,4,5], Shuji Nakamura[1,3,4], Steven P. DenBaars[1,3,4] and Jon. A. Schuller[1].

1. Department of Electrical and Computer Engineering, University of California Santa Barbara
2. Department of Physics, University of California Santa Barbara
3. Department of Material Science and Engineering, University of California Santa Barbara
4. Solid State Lighting and Energy Electronics Center, University of California Santa Barbara

**Table of Contents**

1. Absorption and emission enhancements in structured surfaces
2. Measuring far-field radiation with momentum-resolved photoluminescence spectroscopy
3. Analytical model for far-field radiation from emitting metasurfaces and thin films
4. Complete comparison of experimental and calculated energy-momentum spectra
5. Enhanced and blue-shifted PL due to strain relaxation in InGaN quantum wells

**1. Absorption and emission enhancements in structured surfaces**

In the manuscript, we show an approximately 30-fold increase in the net photoluminescence (PL) intensity from structured surfaces compared to unstructured (thin film) surfaces. This observed enhancement could in general be attributed to both material and optical effects at both the absorption and emission wavelengths. Fig. S1 shows numerical calculations of normalized total absorption enhancements (i.e., relative to absorption in thin films) as a function of pillar width for normal incidence excitation at 405nm (blue). Our metasurfaces are comprised of resonators with widths ranging from approximately 150 nm to 225 nm (shaded region). Within this range, we find an average absorption enhancement of approximately 15. The remainder is attributed to material effects and Purcell emission enhancements, which must therefore be a small contribution (approximately 2). We numerically calculated electric dipole Purcell enhancements according to the description provided in the Methods section of the manuscript. The Purcell factor for emission at 540 nm is presented in Fig. S2 (red) as a function of pillar width for several exemplary pillar widths. Purcell factors in these systems is approximately 1, on average, indicating no significant emission enhancements due to optical resonances alone.

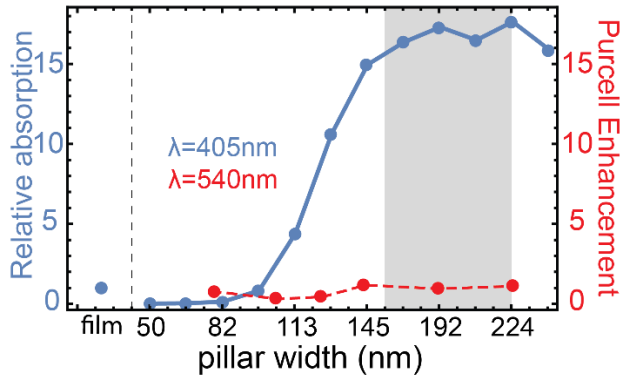

**Figure S1** A) Total absorption in quantum wells in uniform pillar arrays relative to unstructured (film) samples at 405 nm normal incidence excitation is plotted on the left axis. The red curve represents the Purcell enhancement in the QW as a function of the pillar width is plotted w.r.t right axis.

2. **Measuring far-field radiation with momentum-resolved photoluminescence spectroscopy**

Far-field radiation patterns were measured by Fourier microscopy by imaging the back focal plane of an oil immersion objective (numerical aperture NA=1.3)[1-5]. Samples were optically pumped using a 405 nm light emitting diode (LED) (ThorLabs M405L3). The incident light was filtered via a 405 nm short-pass filter and reflected off a 415nm dichroic mirror. The emitted photoluminescence (PL) was transmitted through the same dichroic as well as a 417 nm long-pass filter. The two-dimensional momentum-space ($k_x$, $k_y$) distribution of light was measured by imaging the objective BFP to an imaging spectrometer (Princeton Instruments Iso Plane SCT320 with Princeton Instruments PIXIS 1024BRX). An analyzing polarizer was applied such that light along the $k_x$-axis was either p-(x) or s-(y) polarized. Wavelengths below 525 nm and above 535 nm were eliminated by applying long- and short-pass filters in the collection beam path, respectively. Energy-momentum spectra were measured by applying an entrance slit to the imaging spectrometer. The entrance slit serves to take a narrow line-cut of BFP spectra (e.g., along the $k_x=0$ axis). The light collected through the entrance slit was then spectrally separated.

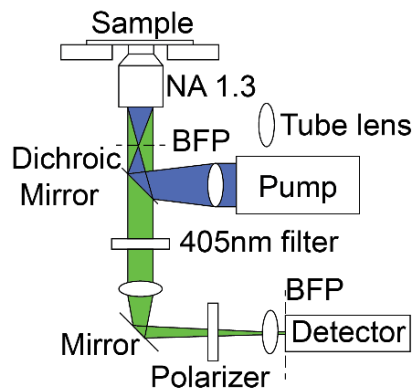

**Figure S2**: Energy-momentum PL measurement setup with a high NA oil immersion lens

\cite{lieb_single-molecule_2004, taminiau_quantifying_2012, schuller_orientation_2013, kurvits_comparative_2015, brown_enhancing_2017}

\cite{lieb_single-molecule_2004, schuller_orientation_2013}.

### 3. Analytical model for far-field radiation from emitting metasurfaces and thin films

Far-field radiation patterns represent the transverse momentum distribution of electromagnetic radiation emitted from a source. In particular, we're interested in the transverse momentum distribution of light, $\rho(\lambda,k_x,k_y)$, of wavelength $\lambda$, emitted into the substrate from a (structured) surface with a growth axis along the z-direction. The quantity $\rho(\lambda,k_x,k_y)$ is known as the local density of optical states (LDOS). In practice, $\rho(\lambda,k_x,k_y)$ is measured by Fourier microscopy (see Sect. S2). Thin film radiation patterns, sourced from oriented electric dipoles, are readily calculated according to the formalism presented in refs. Here, we describe a simple and intuitive analytical theory that may be used to understand unidirectional luminescence observed from our phased array emitting metasurfaces. The theory is developed in three main stages according to the following concept flow:

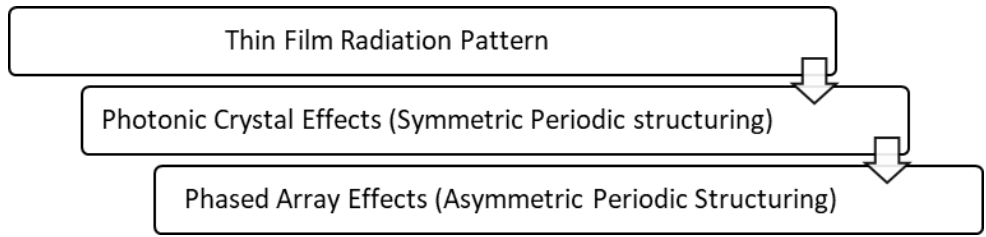

Our emitting metasurfaces begin with InGaN quantum wells embedded in a GaN thin-film. Emission from these quantum wells is characteristic of in-plane emitting electric dipoles [6–9]. For electromagnetic modeling purposes, the essential parameters we need here are the total film thickness (h=1.45µm), the emitting quantum well height (z=1.35 µm), and the GaN refractive index ($n_{GaN}$=2.23, measured at 540 nm). Fig. S3A shows the calculated horizontally (x-) polarized thin-film radiation patterns from this system at $\lambda$ =530 nm, i.e., $\rho^{(x)}(\lambda$ =530 nm; $k_x$, $k_y$). Roughly speaking, the features in this radiation pattern are circular and centered about the origin, {$k_x$, $k_y$}={0,0}. Critically, an in-plane (x-y) oriented electric dipole emitter at a distance 100 nm ($\approx \lambda/2n_{GaN}$) from the GaN-air interface exhibits a natural emission maximum along the $k_y$-axis at $k_y \approx 1.06 k_0$. Note that light along this axis defines p-polarized radiation. Our emitting metasurfaces consist of a periodic square array, with periodicity p=250 nm, of rectangular nanopillars. This "microperiodicity", p, defines a lattice momentum $G_0=2\pi/p$ that symmetrically couples modes with $\vec{k}$ to the harmonics $\{\vec{k} + mG_o\hat{x} + nG_o\hat{y}\}$, for positive or negative integers m and n[6]. This can be described according to traditional photonic crystal (PC) effects, by which we obtain electromagnetic "Brillouin zones". In effect, this serves to "reimage" the thin film LDOS to a grid of images centered at the points {$mG_0$, $nG_0$}. The LDOS of this system is described by

$$\rho^{PC}(\lambda, k_x, k_y) = \sum_{n,m} \rho(\lambda, k_x - mG_o, k_y - nG_o)\Theta\left[n_{sub}k_0 - \sqrt{(k_x - mG_o)^2 + (k_y - nG_o)^2}\right],$$

Where $\Theta(\alpha)$ is the Heaviside step function in the variable $\alpha$, which here serves to ensure that only radiative modes are considered. The calculated $\rho^{PC}(\lambda; k_x, k_y)$ is shown in Fig. S4B. In this calculation, we ignore changes in the effective film refractive index due to the decreased fill fraction (i.e., the removal) of GaN. With the given dimensions and wavelength of interest, $G_0=2.12k_0$. Consequently, the intense emission lobes (at $k_y \approx 1.06k_0$ in the thin-film) between neighboring images nearly perfectly overlap, amplifying the strong emission lobe. Finally, the linear (wrapped) phase gradient, defining our metasurface gratings, defines an asymmetric periodicity, P. This "macroperiocity", P, is associated with a "metasurface momentum" $k_M=2\pi/P$ that contributes an additional asymmetric diffractive component to the emitted light. We define our coordinate system such that our metasurfaces phase gradient is along the x-direction. Thereby, modes with $\vec{k}_o$ are further coupled to the harmonics $\{\vec{k}_o + N\vec{k}_m\}$. In effect, this serves to "translate" the PC LDOS along the $k_x$ axis:

$$\rho^M(\lambda, k_x, k_y) = \rho^{PC}(\lambda, k_x - k_M, k_y)$$

The calculated $\rho^M(\lambda; k_x, k_y)$ is shown in Fig. S3C for a metasurface with $k_M=0.98k_0$ (transmission angle $\theta=80°$ in air). Note that this translation brings the bright emission lobe into the escape cone ($|\vec{k}| < 1$), thus greatly enhancing the light out-coupling efficiency and providing unidirectionality. Note that this whole procedure can be performed just as well for x-polarized radiation profiles. The calculated metasurface radiation profiles show good agreement with experiment. Experimental x-polarized BFP images are shown in Fig. S4 for both a thin-film (Fig. S4D) and the $k_M=0.98k_0$ metasurface (Fig. S1E). These experimental images contain contributions from a wavelength window of approximately $\lambda=525-535$ nm. Consequently, some fine features apparent in the calculations are not evident in the experimental images. Other minor discrepancies between experiment and calculation are attributed to fabrication imperfections and the fact that our model treats the structure as an effective thin-film with homogenized optical constants. Though this is a reasonable approximation due to the sub-wavelength nature of the structuring, it doesn't account for microscopic variations in the spatially-varying effective refractive index of the system. Though, these higher-order corrections could be implemented, the current model offers a simple, succinct, and intuitive description of the observed unidirectional photoluminescence.

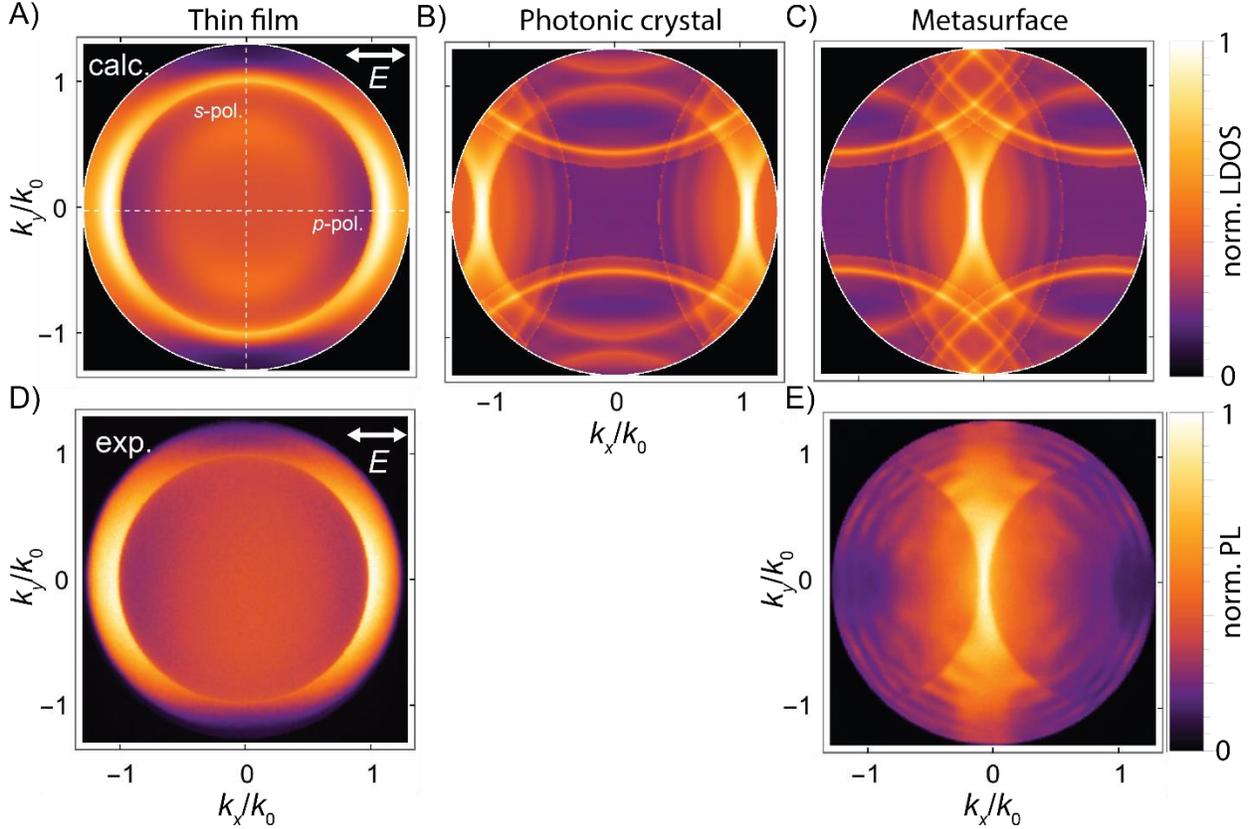

**Figure S3**: Quantitative schematic of calculating metasurface far-field radiation (A-C) Calculated vertically (y-) polarized LDOS at $\lambda_o$=530 nm at three different levels of the theory: (A) thin-film, $\rho(\lambda_0;k_x,k_y)$; (B) photonic crystal ($G_0/k_0$=2.12), $\rho^{PC}(\lambda_0;k_x,k_y)$; (C) metasurface grating ($k_M/k_0$=0.98), $\rho^M(\lambda_0; k_x, k_y)$. (D,E) Experimental x-polarized BFP images of (D) a thinfilm and (E) the $k_M/k_0$=0.98 metasurface. Vertical (horizontal) dashed lines in (A) represent s-(p)-polarized line-cuts

### 4. Comparison of Experimental and Calculated energy-momentum spectra

This model and design principles can be further understood by comparing the calculated and experimental polarized energy-momentum spectra. Fig. S4 shows the (Figs. S4A-D) calculated and (Figs. S4E-H) experimental p-polarized energy-momentum spectra for metasurfaces with $k_M$=(A,E) 0.17; (B,F) 0.5; (C,G) 0.86; (D,H) 0.98. Figs. S4I-P show the corresponding s-polarized (Figs. S4I-L) calculated and (Figs. S4M-P) experimental energy-momentum spectra. Since the normalized lattice momentum, $G_0/k_0 = \lambda/p$, increases linearly with emission wavelength, the emission features disperse linearly with wavelength. With these spectra, we understand the apparent "dip" in emission in s-polarized spectra presented in the manuscript; as $G_0/k_0$ increases beyond approximately 2.08, an emission gap "opens" between neighboring photonic Brillouin zones (see $k_x$-axis of Fig. S4A-B). Curiously, and to great benefit, the "lower" emission lobe (i.e., the bright lobe at $k_y \approx -1.06k_0$ in thin-films) in our metasurfaces becomes dramatically suppressed immediately upon phasing the structure. That is, our structures truly are unidirectional, and this single bright lobe lies within $\overrightarrow{|k|} < 1$. This effect is observed in the experimental p-polarized energy-momentum spectra in the $k_M=0.17$ metasurface (Fig. S4E) as a minimum in emission. Though this feature lies outside the numerical aperture (NA=1.3) of our primary objective for $k_M>0.2$, we have verified its absence using an NA=1.49 objective.

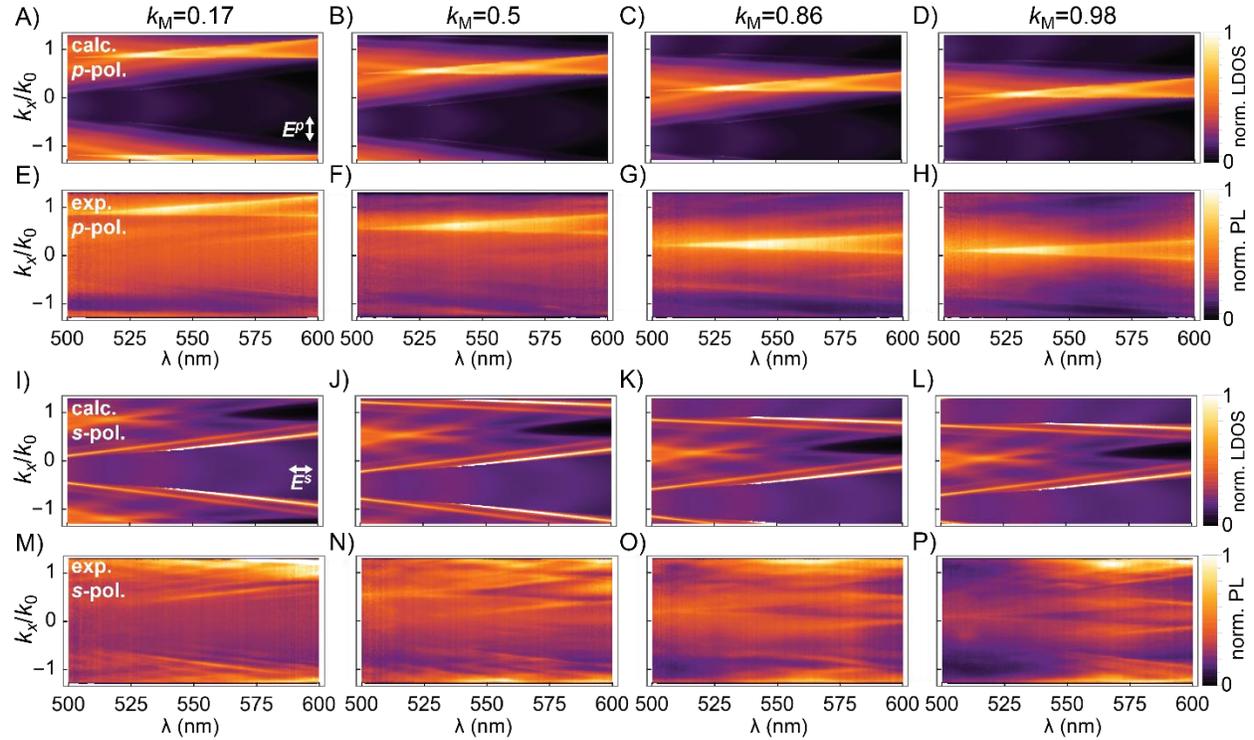

**Figure S4: Comparing calculated and experimental radiation profiles** (**A**-**D**) Calculated p-polarized energy-momentum spectra for metasurfaces with $k_M$=(**A**) 0.17; (**B**) 0.5; (**C**) 0.86; (**D**) 0.98. (**E**-**H**) Experimental p-polarized energy-momentum spectra for metasurfaces with $k_M$=(**E**) 0.17; (**F**) 0.5; (**G**) 0.86; (**H**) 0.98. The corresponding s-polarized (**I**-**L**) calculated and (**M**-**P**) experimental energy-momentum spectra.

### 5. Enhanced and blue-shifted PL due to strain relaxation in InGaN quantum wells

In order to qualitatively explain the blue-shift in PL wavelength of the nanostructures, band structure simulations were carried out using the one-dimensional Drift-diffusion Charge Control solver (1D-DDCC) [7]. The structure used in the simulation is composed by an active zone with three $In_{0.35}Ga_{0.65}N$ quantum wells of 3nm surrounded by 10 nm GaN barrier with a 2nm AlGaN cap on top of each quantum wells. The active zone is in between two unintentionally doped GaN Layer of 100 nm. As previously demonstrated [8-10], InGaN QWs are partially strain relaxed at the center of the nanopillars while they are completely relaxed at the edge. This strain relaxation partially eliminates the built-in electric field in the QWs and reduce the quantum confined stark effect w.r.t the thin film. So, we simulate the band structure of the thin film with the piezo-electric strain while we assume completely strain relaxed InGaN QWs in 150nm GaN nanopillars. In Figure S5B, the simulated band diagram for the 150 nm structure (Plain, Red) and for the continuous film (Dashed, Black) are presented. Figure 1b is a zoom of the active region. The structure with GaN nanopillars presents a larger bandgap together while having a better electron-overlap thanks to the reduced quantum confined Stark effect. We solve the Schrodinger equation to calculate the electron and hole ground state and their wave-function overlap to plot the emission spectrum for both the thin film and the nanopillar structures. From figure S5 A, one can see that the strain-relaxed structure emits a blue-shifted light compared to the continuous film. The shift is in good qualitative agreement with the experimental measurement indicating that the majority of the emission happens close to the strain relaxed surface states in the GaN nano-pillars.

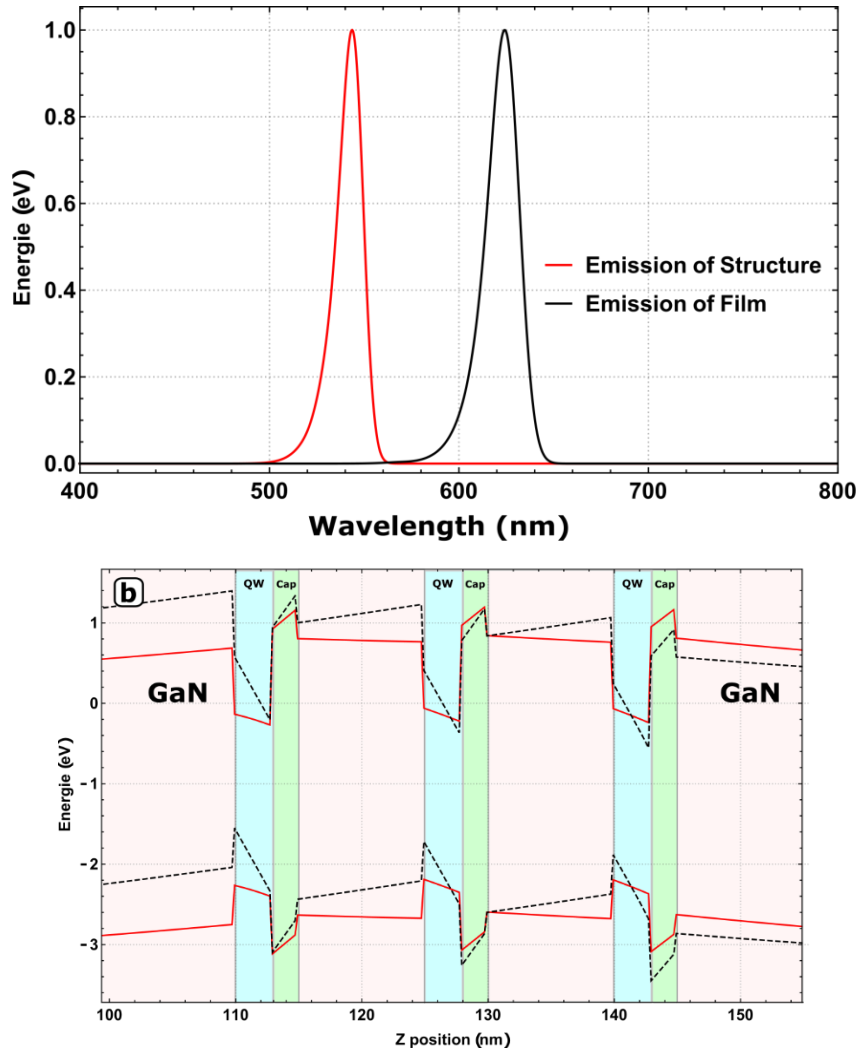

**Figure S5 A)**: Calculated emission spectra: the reduced quantum confined stark effect enables better electron-hole overlap in the structure leading to the blue to shift in the PL spectra. **B)** Simulated band diagram for a 130 nm structure (Red, plain) and for a continuous film (Black, dashed). The 150 nm Pillar-structure is strain relaxed, so the QCSE is less pronounced for the nano-pillars.